\documentclass[aps,superscriptaddress,prl,twocolumn]{revtex4}
\usepackage{epsfig,graphics,amssymb,amsmath,subeqnarray}
\def\p{\partial}
\def\e{{\bf e}}\def\L{L_\perp}\def\E{{\bf E}}\def\u{{\bf u}}
\begin{document}
\title{Hydrodynamic attraction of swimming microorganisms by surfaces}
\author{Allison P. Berke}
\affiliation{
Department of Mathematics, Massachusetts Institute of Technology, 77 Mass. Ave., Cambridge, MA 02139.}
\author{Linda Turner}
\affiliation{
Rowland Institute at Harvard, 100 Edwin H. Land Bld., Cambridge, MA 02142.}
\author{Howard C. Berg}
\affiliation{
Rowland Institute at Harvard, 100 Edwin H. Land Bld., Cambridge, MA 02142.}
\affiliation{
Department of Molecular and Cellular Biology, Harvard University, 16 Divinity Ave.,  Cambridge, MA 02138.}
\author{Eric Lauga\footnote{Corresponding author. Email: elauga@ucsd.edu}}
\affiliation{
Department of Mechanical and Aerospace Engineering, University of California San Diego,\\ 9500 Gillman Drive, La Jolla CA 92093-0411.
}
\date{\today}
\begin{abstract}
Cells swimming in confined environments are attracted by surfaces. We measure the steady-state distribution of smooth-swimming bacteria ({\it Escherichia coli}) between two glass plates.  In agreement with earlier studies,  we find a strong increase of the cell concentration at the boundaries. We demonstrate theoretically that hydrodynamic interactions of the swimming cells with solid  surfaces  lead to  their re-orientation in the direction parallel to the surfaces, as well as their attraction by the closest wall. A model is derived for the steady-state distribution of swimming cells, which compares favorably with our measurements. We exploit our data to estimate the flagellar propulsive force in swimming {\it E. coli}.
\end{abstract}


\maketitle


The majority of swimming microorganisms involved in human functions and diseases are found in geometrically confined environments.  Spermatozoa in the female reproductive tract swim in constricted domains \cite{suarez06}. Bacteria  make their way through host cells and tissues \cite{braybook}, and aggregate in antibiotic-resistant biofilms on surfaces \cite{costerton95}.

Despite the ubiquitous nature of biological motility near surfaces, not much is known about the physical consequences of locomotion in a confined environment \cite{vanloosdrecht90,harshey03}. 
Perhaps the simplest observed  effect of locomotion near walls is the accumulation of swimming cells on  surfaces. In 1963,  Rothschild measured the distribution of bull spermatozoa swimming between two glass plates (separation, 200~$\mu$m). The cell distribution
was  nonuniform, with a constant density in the center strongly increasing near the walls \cite{rothschild63}. Similar results were later obtained for human spermatozoa in glass tubes \cite{winet84a}. 
Further studies for animal spermatozoa pointed out the possible important of  three-dimensional effects  \cite{cosson03,woolley03}.
 Numerical simulations of  model cells with two-dimensional beat patterns \cite{fauci95} supported an explanation in terms of cell-surface hydrodynamic interactions, a scenario   confirmed by recent computations for suspensions of  simplified low-Reynolds number swimmers \cite{hernandez-ortiz05}.
More recent work focused on the change in swimming kinematics near solid walls
\cite{frymier95,diluzio05,magariyama05,lauga06}.

In this paper, we study the attraction of  swimming bacteria by solid surfaces. We measure the distribution of non-tumbling {\it E. coli} \cite{bergbook} cells swimming between two glass plates in a density-matched fluid, and obtain results qualitatively similar to that of Rothschild  \cite{rothschild63}.  We demonstrate theoretically that the origin for the cell profile is  purely hydrodynamical. Using physical arguments based on long-range hydrodynamics interactions between swimming cells and surfaces, we show that these interactions induce  a reorientation of the cells in the direction parallel to the surface, independently of their initial condition (position, orientation), and the subsequent attraction of the cells by the closest wall. Our model allows us to predict the resulting steady-state cell distribution, and is exploited to obtain an estimate for the flagellar propulsive force in swimming {\it E. coli}.


\begin{figure}[b]
\centering
\includegraphics[width=0.42\textwidth]{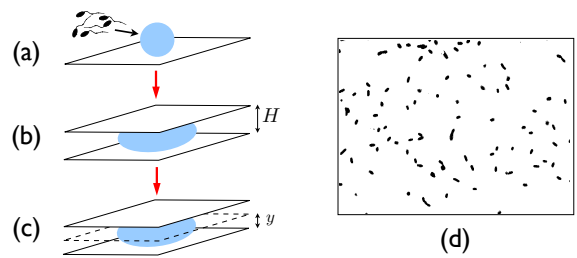}
\caption{Representation of the experimental procedure.
(a): Smooth-swimming {\it E. coli} cells are mixed with a density-matched fluid;
(b): The cell mixture is deposited between two glass plates (separation distance $H$);
(c): The distribution of swimming cells is imaged as a function of the distance, $y$, from the lower surface; 
(d): Example of image obtained from data acquisition in the first layer above the glass surface.
}
\label{procedure}
\end{figure}

\begin{figure}[t]
\centering
\includegraphics[width=0.43\textwidth]{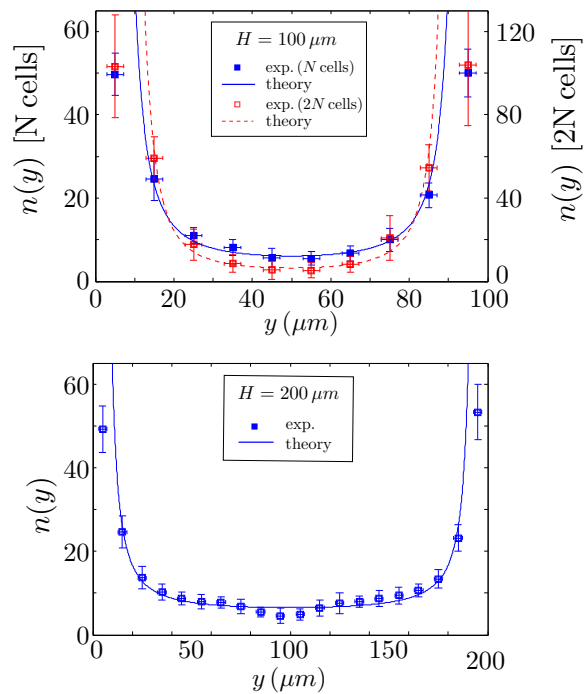}
\caption{
Experimental data: number of swimming cells, $n$, as a function of the distance to the bottom coverslip, $y$, when  the distance between the surfaces is $H=100$ $\mu m$ (top) and $H=200$ $\mu m$ (bottom).
The lines are fit to the  data with the model of Eq.~\eqref{model} with 
$n_0=1.5$ and $L_\perp=34.8$~$\mu$m (top, solid line),
$n_0=0.3$ and $L_\perp=59.1$~$\mu$m (top, dashed line)
and $n_0=3.9$ and $L_\perp=26.4$~$\mu$m (bottom, solid line).}
\label{100200}
\end{figure}

Our experimental procedure is illustrated in Fig.~\ref{procedure}.
{\it E. coli} (smooth-swimming strain HCB-437 \cite{wolfe88}) is grown to mid-exponential phase in T-broth (1\% Tryptone, 0.5\% NaCl), washed three times by centrifugation (2200g for 8 min), and then resuspended in a motility medium (10 mM potassium phosphate, pH 7.0, 0.1 mM EDTA).  
PVP-40 (polyvinylpyrrolidone) is added (0.005\%) to prevent adsorption of cells to glass, and the final suspension is combined with Percoll (2:3 ratio) to match the medium and cell buoyant densities \footnote{Matching of buoyant densities allows gravity to be irrelevant, as further demonstrated by the symmetric profiles in Fig.~\ref{100200}.}. A droplet of the cell mixture is deposited between two glass coverslips, previously cleaned in a mixture of ethanol saturated with potassium hydroxide, rinsed with
ultra-pure filtered water, and allowed to air dry. The coverslips are separated by a distance $H$, controlled by layers of other coverslips (\#1.5) and verified by caliper measurement. 
A phase-contrast microscope (Nikon Optiphot-2) using 600x
magnification (depth of field, $4.3$ $\mu$m) and equipped with a shuttered CCD video camera (Marshall Electronics V1070) set for an exposure of 1 ms/frame is used to image the population of swimming cells. 
The video signal is sent to a MacG4 equipped with an LG-3 frame grabber (Scion Image Corp.) and ImageJ software (NIH, Bethesda, MD).  We capture 2-second
movies at 20 frames per second and measure the number
of swimming cells by counting cells swimming at speed larger than 1 body length per second. We start 5~$\mu$m above the lower glass surface; we then bring the plane of focus up 10~$\mu$m, and repeat the measurement until we reach within 5~$\mu$m of the upper glass surface. Experiments are then repeated with other cell samples and  sets of coverslips.

In our protocol,  two parameters can be varied:  the distance, $H$, between the two coverslips (we chose $H=100$~$\mu$m or $200$~$\mu$m) and the cell density of the final mixture, {\it i.e.}, the size of the overall cell population (when $H=100$~$\mu$m, we performed additional experiments doubling the number of cells).
The experimental results are shown in Fig.~\ref{100200};  vertical errors bars represent statistics on ten different experiments, and horizontal error bars the  depth of field. As in Ref.~\cite{rothschild63}, we find that the cell profile peaks strongly near the walls, with a nearly constant cell density about 20~$\mu$m away from the walls; this is the main experimental result of this paper. 

We now turn to the physical understanding of the attraction phenomenon. In order to provide a complete physical picture, we need to identify  the mechanism responsible for the non-uniform cell distribution, and predict the steady-state profile observed experimentally.

\begin{figure}[t]
\centering
\includegraphics[width=.49\textwidth]{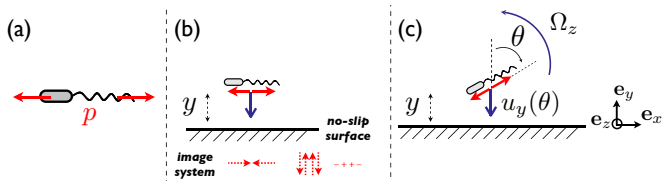}
\caption{Attraction of microorganisms to solid surfaces. 
(a) The flow field around a swimming cell is well approximated by a force-dipole of strength $p>0$, represented by two arrows pointing in opposite directions;
(b) Hydrodynamic attraction of a force-dipole by a no-slip surface due to the image system on the other side of the surface (force-dipole, force-quadrupole,  source-quadrupole);
(c) Notations for the model.}
\label{attraction}
\end{figure}

The physical mechanism for the attraction is the hydrodynamic interactions between swimming cells and surfaces \cite{fauci95,hernandez-ortiz05}. The flow around  most flagellated swimming organisms, including  spermatozoa cells or bacteria such as {\it E. coli}, is well approximated by a force-dipole (stresslet) \cite{pedley92}: the  flagellar motion provides the propulsive force which is opposed by the drag on both the cell body and flagella, corresponding to a force-dipole in which both the flagella and the body act on the fluid in the direction away from the cell (represented in  Fig.~\ref{attraction}a by two arrows pointing in opposite directions). The fluid velocity is given by 
${\bf u}=\frac{p}{8\pi\eta r^3}\left(-1+3\frac{({\bf r}\cdot \e)^2}{r^2} \right){\bf r}$, where $p>0$ is the dipole strength, $\e$  the swimming direction, $\eta$ the viscosity, and ${\bf r}$ the distance to the dipole; this far-field model is valid for distances larger than the length, $L$, of the swimming cells (body plus flagella), an approximation that we will make in this paper.

Near a wall, the flow field  induced by the cell is a superposition of that due to the force-dipole, plus any image flow field, located on the other side of the surface, and necessary to enforce the correct surface boundary condition (similar to the method of images in electrostatics, only here the image is a vector field). 
The image system for a force-dipole parallel to a no-slip surface is known \cite{blake71}  (force-dipole, force-quadrupole and source-quadrupole) and is found to induce, at the location of the dipole, a velocity component towards the solid surface of order $  \sim p/\eta y^2$ where $y$ the distance to the surface (Fig.~\ref{attraction}b); this wall-induced flow is the reason for the attraction  \cite{hernandez-ortiz05}. 
To gain physical intuition, it is easier to picture a dipole near a free surface; in that case the image system is an equal dipole on the other side of the surface, and two parallel dipoles attract each other. Physically, this wall-induced velocity is reminiscent of the shear-induced migration of deformable bodies and polymers  away from boundaries  \cite{leal80,agarwal94,seifert99,cantat99}, with the difference that in these cases, the  dipoles arise from  shear-induced deformation, and therefore they have the opposite signs ($p<0$).
 
Although the aforementioned dipole-dipole attraction contains the essential physical picture, some further considerations are required. When the force-dipole is not aligned with the nearby surface, the attraction can become a repulsion. Indeed, if we denote by $\theta$ the orientation of the dipole w.r.t. the vertical direction, the induced velocity in the direction away from the no-slip wall is given by (Fig.~\ref{attraction}c)
\begin{equation}
\label{uy}
u_y (\theta,y)= -\frac{3p}{64\pi \eta y^2}(1-3\cos^2\theta).
\end{equation}
Notably, if the angle is small enough ($\theta < \arccos 1/\sqrt{3}$), the induced velocity changes sign and turns into a repulsion ($u_y>0$) \footnote{For a free-surface, $u_y (\theta,y)  =   - p (1-3\cos^2\theta)/(32\pi \eta y^2)$.}. Moreover, the average of the wall-induced attraction over a whole population of randomly oriented microorganisms is exactly equal to zero, $\int u_y \sin\theta\, {\rm d}\theta = 0 $. Consequently, any asymmetry in the final distribution of swimming cells due to this mechanism alone will reflect  the non-uniformity of the initial distribution of cell orientations, and is therefore expected to be small for large populations.

Here, we propose that hydrodynamic interactions with the surfaces provide the physical mechanism to modify the orientations of the cells. Let us consider the complete flow field due to a swimming cell near a solid surface. Not only does the image system for the force-dipole induces a local attractive/repulsive velocity for the cells, it also contains (in general) non-zero velocity gradients; these gradients are responsible for the rotation of the cells. Modeling the swimming cells as force-free and torque-free prolate spheroids of aspect ratio $\gamma$, their rotation rate, $\boldsymbol{\Omega}$, is given by 
\begin{equation}
\label{Omega}
\boldsymbol{\Omega} = \frac{1}{2} \boldsymbol{\omega} +\left(\frac{\gamma^2-1}{\gamma^2+1}\right) \e \times (\E \cdot \e),
\end{equation}
where $\boldsymbol{\omega} =\nabla \times \u$ and $\E=\frac{1}{2}(\nabla \u + \nabla \u^T)$ denote, respectively,  the vorticity and the rate-of-strain of the flow field due to the image system \cite{kimbook}.
Eq.~\eqref{Omega} states that the cells are rotating at one half the local value of the vorticity generated by the image system, plus an additional term that depends on the aspect ratio of the cell and the straining flow component of the image velocity field.

Using the notations defined in Fig.~\ref{attraction}, we evaluate the component of the rotation rate in the direction $z$, parallel to the surface and perpendicular to the dipole, and obtain
\begin{equation}\label{Oz}
\Omega_z (\theta,y)=  -\frac{3 p \cos\theta \sin\theta}{64 \pi \eta y^3} 
\left [ 1 + \frac{(\gamma^2-1)}{2(\gamma^2+1)}(1+ \cos^2\theta)\right].
\end{equation}
Since,  for {\it E. coli},  $p>0$ and  $\gamma \gg 1$
($\gamma\approx 2$ for the cell body, and $\gamma \gg 1$ for the flagellar bundle),
 Eq.~\eqref{Oz} shows that $\Omega_z $ has always the same sign as $-\cos\theta \sin\theta$. 
When $0\leq \theta \leq \pi/2$, the rotation rate is negative, and brings the cell in the direction parallel to the surface; 
when $\pi/2\leq \theta \leq \pi$, the rotation rate is positive, and  also aligns the swimming cell  parallel to the surface. 
Consequently all swimmers are reoriented in the direction parallel to the surface by hydrodynamic interactions \footnote{For a free-surface, we have $\Omega_z (\theta,y)  =   -3 p \cos\theta \sin\theta 
\left ( 1 + \frac{\gamma^2-1}{\gamma^2+1}\cos^2\theta\right)/(64 \pi \eta y^3)$.}. 

On what time scale do the cells reorient? Cells  not initially parallel to the surfaces first swim towards one surface, on a time scale $\tau_U\sim H/U$, where $U$ is their swimming speed. When they reach a distance $y\sim L$ from the surface, reorientation takes place on a time scale  
$\tau_\Omega\sim \Omega_z^{-1}\sim \eta L ^3 / p$. 
From a scaling standpoint, the dipole strength, $p$, is on the order of the drag (or thrust) force on the organism times the typical cell size, $p\sim \eta U L^2$, and therefore we have $\tau_\Omega\sim L/U$. 
For {\it E. coli} cells of size $L\approx 5-10$ $\mu$m
swimming at $U\approx 20$ $\mu$m s$^{-1}$   \cite{chattopadhyay06,darnton07}, the reorientation occurs in  a matter of seconds. By comparison, the time scale  $\tau_R$ for reorientation by rotational Brownian motion is  $\tau_R\sim D_R^{-1}$, where $D_R$ is the rotational diffusivity, $D_R\sim k_B T/\eta L^3$ ($k_B$ is Boltzmann's constant and $T$ the temperature). For {\it E. coli} cells, we expect  $\tau_R\sim 10^2$ s, and therefore the swimming directions of the cells are dominated by the (deterministic) hydrodynamic mechanism outlined above.

With cells swimming parallel to the surfaces, the steady-state cell probability distribution across the chamber, $n(y)$, is described by a balance between  the advection by the  velocity field in Eq.\eqref{uy} and cell diffusion (with diffusivity, $D_\perp$), so the cell conservation equation is written as 
\begin{equation}\label{advdiff}
\frac{\p}{\p y}\left(n u_y\right) =  D_\perp \frac{\p^2 n}{\p y^2}\cdot
\end{equation}
For cells parallel to two surfaces separated by a distance $H$, the attractive velocity is written as
\begin{equation}
\label{2walls}
u_y (0,y) = -  \frac{3p}{64\pi \eta }\left(\frac{1}{y^2}-\frac{1}{(H-y)^2}\right),
\end{equation}
plus terms of higher order in $1/y$ and $1/H$, which are due to reflections of each image system on the opposite surfaces and are neglected \cite{liron76,kimbook}.  Integration of Eq.~\eqref{advdiff} using Eq.~\eqref{2walls} and assuming that $\p n /\p y=0$ halfway between the two plates leads to an analytical model for  the steady-state concentration profile as
\begin{equation}\label{model}
\frac{n(y)}{n_0} =  \exp\left[L_\perp\left( \frac{1}{y}+\frac{1}{H-y}\right)\right],\,\,\,\,
L_\perp = \frac{3p}{64 \pi \eta D_\perp}\cdot
\end{equation}

The comparison between our data and this model are presented in Fig.~\ref{100200}. 
We fit the two parameters in Eq.~\eqref{model}, $n_0$ and $\L$,  to the data points located further away from the closest wall than $h \sim L\approx 10$~$\mu$m, below which our point-dipole model is no longer valid \footnote{The depth of field is much smaller than both $H$ and $L_\perp$, so we can directly compare the model from Eq.~\eqref{model} with our experimental results}. The agreement between the theory and our experimental results is  good.  
Close to the wall, we have $(y,H-y)\ll L_\perp$, and the model over-predicts the cell density. 
This could be regularized by modeling near-wall hydrodynamics \cite{ramia93,lauga06}, and including other cell-wall interactions (intermolecular and screened electrostatics), but would not modify our far-field results.
For the data with $H=100$~$\mu$m, the best fits with our model (in a least-square sense) are obtained for $\L=48.3\pm 15$~$\mu$m; 
when $H=200$~$\mu$m, agreement is obtained for 
$\L= 26.9\pm 4$~$\mu$m.

We now exploit our results to estimate the dipole, $p$,  for smooth-swimming {\it E. coli} cells. From Eq.~\eqref{model}, we see that  for a fluid of known viscosity, $\eta$, one only needs the value $D_\perp$ and the estimate of $\L$,  to obtain the estimate $p= 64 \pi \eta D_\perp \L /3 $. 
Fitting the measurements to our model leads to  $ \L=20-76$~$\mu$m.
Since the cells swim parallel to the surfaces, the diffusion coefficient, $D_\perp$, is the Brownian diffusivity for bacteria in the direction perpendicular to their swimming direction \footnote{For larger cell concentrations, we expect cell-cell hydrodynamic interactions to contribute to $D_\perp$ as well.}; in water \footnote{The product $\eta D_\perp$ is independent of the fluid viscosity and can be evaluated for water.} at 20$^\circ$C, we estimate $D_\perp\approx 0.1-0.2$~$\mu$m$^2$/s, leading to the estimate  $p\approx 0.1-1$~pN\,$\mu$m.  The distance, $\ell$,  between drag- and thrust-producing units on the swimming cell is on the order of the size of the cell body, which displaces the center of drag ahead of the center of thrust, and $\ell \approx 1$~$\mu$m. We therefore estimate from our measurements a thrust force, $f=p/\ell\approx 0.1-1$~pN. This agrees with recent experiments for swimming {\it E. coli} reporting flagella thrust forces $f\approx 0.57$~pN \cite{chattopadhyay06}  and $f \approx 0.41$~pN \cite{darnton07}.


In conclusion, we have studied experimentally and theoretically the attraction of swimming micro-organisms by surfaces. Experimental data with smooth-swimming bacteria display a strong attraction by solid surfaces, which we have rationalized as follows: hydrodynamic interactions with surfaces result in a  reorientation of the swimming cells in the direction parallel to the surfaces, and  an attraction of the aligned cells  by the nearest wall. We have also shown how to exploit the measurement of steady-state population profile to estimate the flagellar propulsive force of the swimming cells, a simple method to estimate motile propulsive forces. As an extension, we note that some swimming cells,
such as the algae {\it Chlamydomonas}, are not pushed from the back by their flagella  as in {\it E. coli}  but are instead pulled from the front. In that case, the sign of the dipole is reversed, $p<0$: the wall-induced rotation rate (Eq.~\ref{Oz}) changes sign, and hydrodynamic interactions reorient the cells in the direction perpendicular to the surface.  For these cells, hydrodynamic attraction is therefore also expected, but for a different reason:  they simply crash into the walls.

This research was funded by 
the NSF (grant CTS-0624830 to EL), 
the Charles Reed Fund at MIT (EL), 
and by the NIH (grant AI065540 to HCB).

\bibliographystyle{prsty}
\bibliography{attraction_surfaces}

\end{document}